\documentclass[preprint]{aastex6}
\usepackage{exscale}
\usepackage{amsmath}
\usepackage{color}
\usepackage{rotating}
\usepackage{graphicx}
\usepackage{epstopdf}
\usepackage{amsfonts}
\usepackage{amssymb}
\usepackage{hyperref}
\usepackage{natbib}
\usepackage{caption2}
\usepackage{subfigure}
\usepackage{overpic}
\usepackage{ulem}
\usepackage{multirow}
\usepackage{array}
\usepackage{booktabs}
\usepackage{appendix}
\usepackage{threeparttable,chngpage,diagbox}
\usepackage{lineno}

\newcommand{\rsun}{R$_{\odot}$}
\newcommand{\kms}{km~s$^{-1}$}

\shorttitle{Is there a Dynamic Difference between Stealthy and Standard CMEs?}
\shortauthors{Ying et al.}

\begin{document}

\title{Is there a Dynamic Difference between Stealthy \\
and Standard Coronal Mass Ejections?}
\author{Beili Ying\altaffilmark{1},Alessandro Bemporad\altaffilmark{2}, Li Feng\altaffilmark{1,3}, Nariaki V. Nitta\altaffilmark{4} and Weiqun Gan\altaffilmark{1,3}}

\email{alessandro.bemporad@inaf.it; lfeng@pmo.ac.cn}

\altaffiltext{1}{Key Laboratory of Dark Matter and Space Astronomy, Purple Mountain Observatory, Chinese Academy of Sciences,210023 Nanjing, China}
\altaffiltext{2}{INAF-Turin Astrophysical Observatory, via Osservatorio 20, 10025 Pino Torinese (TO), Italy}
\altaffiltext{3}{School of Astronomy and Space Science, University of Science and Technology of China, Hefei, Anhui 230026, People's Republic of China}
\altaffiltext{4}{Lockheed Martin Solar and Astrophysics Laboratory, Department A021S, Building 252, 3251 Hanover Street, Palo Alto, CA 94304, USA}
\begin{abstract}
Stealthy Coronal Mass Ejections (CMEs), lacking low coronal signatures, may result in significant geomagnetic storms. However, the mechanism of stealthy CMEs is still highly debated. In this work, we investigate whether there are differences between the stealthy and standard CMEs in terms of their dynamic behaviors. Seven stealthy and eight standard CMEs with slow speeds are selected. We calculate two-dimensional speed distributions of CMEs based on the cross-correlation method, rather than the unidimensional speed, and further obtain more accurate distributions and evolution of CME mechanical energies. Then we derive the CME driving powers and correlate them with CME parameters (total mass, average speed, and acceleration) for standard and stealthy CMEs. Besides, we study the forces that drive CMEs, namely, the Lorentz force, gravitational force, and drag force due to the ambient solar wind near the Sun. The results reveal that both the standard and stealthy CMEs are propelled by the combined action of those forces in the inner corona. The drag force and gravitational force are comparable with the Lorentz force. However, the impact of the drag and Lorentz forces on the global evolution of the stealthy CMEs is significantly weaker than that of the standard CMEs. 
\end{abstract}


\section{Introduction}

\bibliographystyle{apj}

Coronal mass ejections (CMEs) are one of the most violent explosions in the solar atmosphere, which often encompass a wide range of signatures in their life cycle. A typical CME owns a three-part structure, including a bright front, a dark cavity, and a dense core in the white-light coronagraph \citep{Illing1983}. In recent years, the term ``stealth CME" has been mentioned frequently. This term is generally used to describe CMEs 
that lack corresponding eruptive features in the low corona, e.g., filament, flare, post-eruptive arcade, coronal wave, coronal dimming, jet, and so on \citep{Robbrecht2009, Ma2010, Vourlidas2011, HowardTA2013, D'Huys2014}. The stealth CMEs tend to be slow \citep[$<$500~\kms, ][]{D'Huys2014} and their possible source regions are more frequently located near coronal holes or open field regions \citep{Nitta2017}.
Despite low speeds, stealth CMEs can be the sources of quite intense geomagnetic storms  \citep{Cane2003, Zhang2007, Kilpua2014, Nitta2017}, which may belong to the so-called ``problem geomagnetic storms'' \citep[e.g.,][]{McAllister1996} because of the lack of clear solar associations.  Without knowing the origin of the CME, it is extremely difficult to forecast the strong southward magnetic fields of interplanetary CMEs near the Earth, which are pivotal factors in determining the intensities of geomagnetic storms \citep{Tsurutani1992, Nitta2017}. Encouraging results were recently presented by \citet{Palmerio2021ApJ}, who demonstrated that it would be possible to reproduce the arrival and magnetic properties of a stealth CME as observed in situ by Parker Solar Probe, using a new tool that combined data analysis and numerical modelling. 

However, the mechanism of stealth CMEs is still one of the remaining open scientific questions. First of all, how special are stealth CMEs?  \citet{HowardTA2013} proposed that there may be no essential difference in physical mechanisms responsible for stealth CMEs from those for normal CMEs, but that the failure to observe the low coronal signatures of the CME may be due to the limitation of the observing instruments. Advanced image processing techniques \citep[e.g.,][]{Morgan2014} were used to revisit the stealth CMEs as compiled by \citet{D'Huys2014}, revealing some low coronal features off the limb \citep{Alzate2017}. However, these image processing techniques may not detect new on-disk signatures of stealth CMEs that are not found using traditional methods of difference images \citep{Palmerio2021}. As reviewed by \citet{Nitta2021}, there have been some promising developments in modeling stealth CMEs \citep[e.g.,][]{Lynch2016, Talpeanu2020, Yardley2021}. In addition, data of stealth CMEs from new missions are being analyzed \citep[e.g.,][]{O'KaneSolO}. However, it is still premature, at present, to determine whether stealth CMEs are triggered by mechanisms different from those that constitute the standard model for eruptive flares \citep{Svestka1992}. 

Apart from the concern of stealth CMEs' origins in the solar corona, we should not ignore the implications of stealth CMEs in space weather operations. Understanding the dynamics of CMEs and predicting their speeds and transit times from the Sun to the Earth play significant roles in space weather forecasting \citep{Durand-Manterola2017}. Studies in CME dynamics generally focus on forces at work during their propagation from the solar corona into the interplanetary space \citep{Byrne2010,Carley2012,ShenFang2012,Talpeanu2022}. In the propagation stage, CMEs are thought to be driven by the mixture of the internal Lorentz force and the aerodynamic drag force due to the interaction with the ambient solar wind \citep{Byrne2010,Carley2012,Sachdeva2017}. \citet{Subramanian2007} examined a sample of well-observed CMEs and derived their corresponding driving powers based on the temporal evolution of the mechanical (kinetic $+$ potential) energies. They found that the driving power of these CMEs mainly comes from the contribution of the internal magnetic energy without the effect of the solar wind throughout the field of view (FoV) of the Large Angle and Spectrometric Coronagraph \citep[LASCO;][]{Brueckner1995}. Through the analysis of 38 CMEs based on a Graduated Cylindrical Shell (GCS) model \citep{Thernisien2006,Thernisien2009,Thernisien2011}, which is a geometrical model that approximates the CME as two cones attached to the Sun, connected by a bent cylinder at the top, \citet{Sachdeva2017} discovered that the Lorentz force of all CMEs peaks between 1.65 and 2.45 $\rm R_{\odot}$, and it only becomes negligible compared to the drag force at a distance between 12 to 50 $\rm R_{\odot}$ for CMEs slower than 900~\rm km~s$^{-1}$.

Through statistical or case studies, several works analyzed the one-dimensional kinematics (e.g., velocity, acceleration) of stealth CMEs via tracking bright structures of CMEs \citep{Ma2010, D'Huys2014, O'Kane2019, Nitta2021}. Nevertheless, the anisotropy of the CME plasma shows that it is far from enough to use the average velocity of the CME front or core to characterize the CME's kinematic properties. \citet{Feng2015a} estimated the radial flow speed and mechanical energy inside a CME via the radial mass transport process, and found that the kinetic energy is apparently overestimated if it is derived using the total mass and the leading edge motion. 

In this work, we attempt to explore whether there are differences in the driving power and force of two different types of CMEs, standard (normal) and stealthy. We use the cross-correlation method (CCM) that has been developed to measure two-dimensional (2D) radial speeds of CMEs \citep{Ying2019} as observed by space-based coronagraphs. Based on the CCM, we can obtain more accurate kinematics of whole CMEs, and analyze the kinematic properties of different parts of CMEs, such as the core and front. In this work, we select two types of CMEs with slow speeds ($<600~\rm km~s^{-1}$ at the fronts), standard and stealthy events, and apply the CCM to their coronagraphic observations. Detailed definitions of these two types of CME and the data analyzed in this work will be described in Section~\ref{sec:data}. The CME parameters, including the 2D radial speed, mass and energy distributions, are measured in Section~\ref{sec:para}. Then, we derive the driving powers of CMEs, equal to the mechanical (kinetic plus potential) energy rates, to investigate how the CME powers correlate with the CME parameters (total mass, average speed, and acceleration) in stealthy and standard CMEs. The results obtained from the CCM allows us to separate the CME into the core and front and to study how the above correlations differ in the different parts of the CME. This is discussed in Section~\ref{sec:result}. In Section~\ref{sec:force} we calculate the drag force, gravitational force, and Lorentz force acting on the CMEs to compare their behaviors in these two types of CMEs. Conclusions and discussions are given in Section~\ref{sec:conclusion}.

\section{Data}~\label{sec:data}
\begin{table}[]
  \centering
  \caption{List of CME events}
  \label{tab:01}
  \begin{adjustwidth}{-5 em}{0 em}
  \resizebox{1.1\textwidth}{!}{%
  \begin{tabular}{cccccccccccc}
  \hline
  1 & 2 & 3 & 4 & 5 & 6 & 7  & 8 &9 &10 & 11 &12\\
  No. & Date & Time (UT) & PA (STA) & PA (STB) & Latitude & Deviation angle & Core & Fraction & Domain & Source region & Data used \\ \hline 
  \multicolumn{10}{c}{Standard CME} \\ \hline
  1 & 2011 Aug 04 & 04:00 & 295 & 60 & N25 & 13 &- &- & far-side & STB(E70-E85,N13-N32) & STB \\
  2 & 2011 Sep 13 & 21:00 & 60 & 300 & N22 & 2&Y & 0.07 & front-side &AIA(E06-W18,N10-N25) &STA \\
  3 & 2011 Sep 29 & 17:30 & 290 & 45 & N25 & 25&Y & 0.06 & far-side &STB(E45-E65,N20-N40) &STA \\
  4 & 2011 Sep 29 & 20:00 & 320 & 45 & N41 & 20&Y & 0.045 & far-side &STA(W30-W60,N20-N35) &STB \\
  5 & 2011 Oct 28 & 18:00 & 70 & 290 & N21 & 2&- & - & front-side&AIA(E07-W10,N01-N12) &STB \\
  6 & 2012 Jan 03 & 23:00 & 260 & 100 & S11 & 12& Y &  0.05 & far-side &STB(E60-E75,S26-S07) &STB \\
  7 & 2012 Mar 21 & 18:30 & 125 & 235 & S31 & 14&- &  - & front-side &AIA(E60-E15,S20-S60) &STB \\
  8 & 2012 Oct 04 & 23:00 & 115 & 235 & S23 & 20&Y & 0.07 & front-side &AIA(W05-W45,S10-S40) &STA \\ \hline
  \multicolumn{10}{c}{Stealthy CME} \\ \hline
  9 & 2009 Jun 13 & 00:05 & 100 & 280 &S04& 25&Y & 0.07 & front-side &- &STA \\
  10 & 2010 Jun 16 & 07:00 & 100 & 270 &S06& 4&Y & 0.03 & front-side &- &STA \\
  11 & 2010 Dec 22 & 23:00 & 115 & 240 &S15& 12&- &  - & front-side &- &STA \\
  12 & 2011 Jan 19 & 00:20 & 100 & 260 &S04& 10&- & - & front-side &- &STA \\
  13 & 2011 Mar 02 & 15:00 & 110 & 250 &S22& 15&- & - & front-side &- &STA \\
  14 & 2011 Mar 24 & 18:00 & 90 & 270 &S05& 14&Y &  0.07 & front-side &- &STB \\
  15 & 2013 Jun 29 & 23:00 & 70 & 270 &S06 &33&Y &  0.07 & front-side &- &STA \\
   \hline
  \end{tabular}%
  
  }
  \end{adjustwidth}
  \begin{tablenotes}
    \item[1] Column 1: Event number. Column 2: Observation date. Column 3: Start time of the erupted CME appearing in the FoV of COR1. Columns 4 \& 5: Central position angles (PAs, counterclockwise from the solar north in degrees) of CMEs in the FoV of STA and STB, respectively. Column 6: The center latitude of the CME measured by the GCS model. The angle is given in a degree and increases from the solar equator to the solar polar. Column 7: The deviation angles of CMEs away from the PoS of the observing instruments in degrees measured by the GCS model. Column 8: The CME owns clear core structures marked by ``Y". The symbol ``-'' means the CME has no clear core. Column 9: Optimized fractions used to select the signals of the CME cores. Column 10: Front-side or far-side domain of the CME source region. Column 11: Source regions of standard CMEs in heliographic coordinates. We do not list the source region of stealthy CMEs due to their ambiguity. Column 12: The instrument of the data used for analysis in this work.
  
  \end{tablenotes}
  \end{table}


\begin{figure*}[htb]
  \centering
  \includegraphics[width=0.7\textwidth]{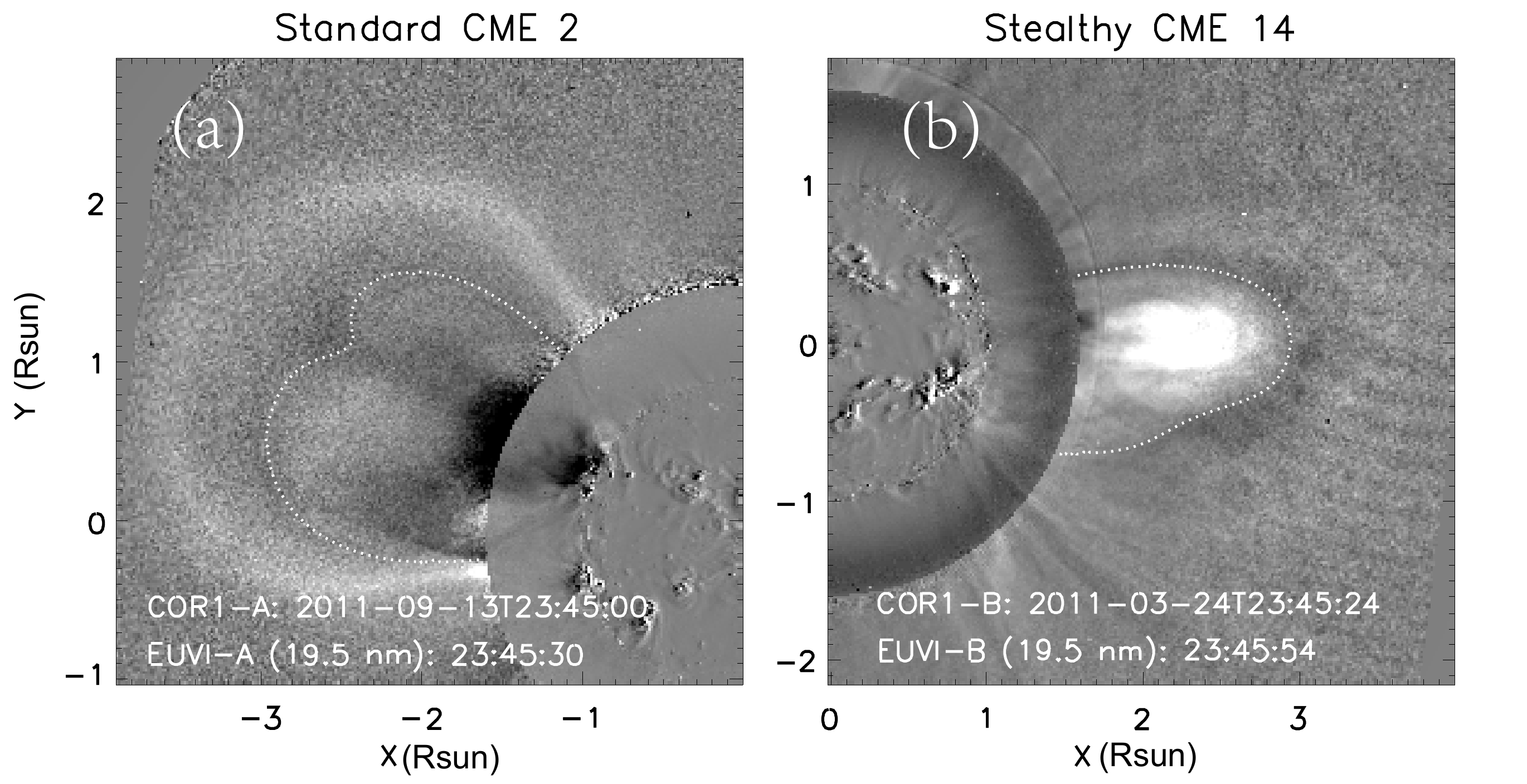}
  
  \caption{Examples of standard and stealthy CMEs. (a): A composite base-difference image of a standard CME (CME 2) from COR1 and EUVI on STA that occurred on 2011 September 13. The preevent background image acquired at $\sim$20:00 UT was subtracted. (b): A composite base-difference image of a stealthy CME (CME 14) occurred on 2011 March 24. The subtracting preevent background was at $\sim$18:00 UT. Dotted lines encircle core regions of CMEs 2 and 14.}
  \label{fig:eventimg}
\end{figure*}

In this work, we select two types of CMEs to compare their dynamic properties. We will analyze the kinematics of CMEs within the FoV of COR1 coronagraphs on board the Solar TErrestrial RElations Observatory (STEREO) mission \citep{Kaiser2008}. To minimize the measurement error caused by the projection effect in the plane-of-sky (PoS), we choose the eruptions as close to the solar limb as possible. Besides, the speeds of CMEs should not be too fast so that we have sufficient data in the FoV of COR1. The front velocities of all CMEs in this work are less than 600 $\rm km~s^{-1}$ in the FoV of COR1.

For the first type of events we call standard CMEs, the selection criteria include: (1) All events are linked to clear low coronal features, such as coronal dimming, flare ribbons, post-flare arcades, and so on. (2) During the CME eruptions, no prominences or filaments erupt into the FoV of STEREO/COR1, to avoid strong H$\alpha$ contributions from prominences to the Thomson-scattered signals in COR1 observations \citep{Mierla2011, Howard2015a, Howard2015b}. (3) These events can be observed by STEREO-A (STA) and STEREO-B (STB) simultaneously. The selected standard CMEs are listed in \autoref{tab:01}. It is noted that four CMEs (1, 3, 4, 6) are far-side events, of which the source regions are undetectable from the Earth's perspective.

The second type of event is called stealthy CME. It is hard to observe corresponding source regions of these events without employing non-standard techniques such as long-time differencing in both EUV and coronagraphs images \citep{Nitta2021}. Thus, similar to \citet{Nitta2017} and \citet{Nitta2021}, we consider CMEs as stealthy when their low coronal signatures are significantly weaker than in standard CMEs. All seven of our stealthy CMEs have been recognized as such in previous studies \citep{Ma2010, Nitta2017}. Of these selected stealthy CMEs, six CMEs (CMEs 10-15 in \autoref{tab:01}) are from \citet{Nitta2017}, and one CME (CME 9) is from \citet{Ma2010}. It is noted that the observation time in \autoref{tab:01} (column 3) is the start time of the CME appearing in the FoV of COR1. 

Using the GCS model, we find that the center latitude of almost all events is between S30$\arcdeg$-N30$\arcdeg$, except for CME 4 ($\sim \rm N41\arcdeg$, column 6 in \autoref{tab:01}). We find dense streamers for most events, including CME 4, before the eruption. The CME latitude range and the existing streamers suggest that all our CMEs were ejected into the slow solar wind. Besides, we determine that almost all CMEs are directed away from the PoS by less than 25$\arcdeg$ in observed perspectives based on the GCS model, except for CME 15, which deviates from the PoS by an angle of about 33$\arcdeg$ (column 7 in \autoref{tab:01}). The angle of CME 15 may cause an error of about 40\% in estimating the kinetic energy. Examples of a standard CME and a stealthy CME are shown in \autoref{fig:eventimg}.

\section{CME parameters} ~\label{sec:para}
This section carries out the corresponding processes on images and data to obtain useful CME parameters, including the 2D velocity, mass, and energy distributions.

\begin{figure*}[htb]
  \centering
  \includegraphics[width=0.7\textwidth]{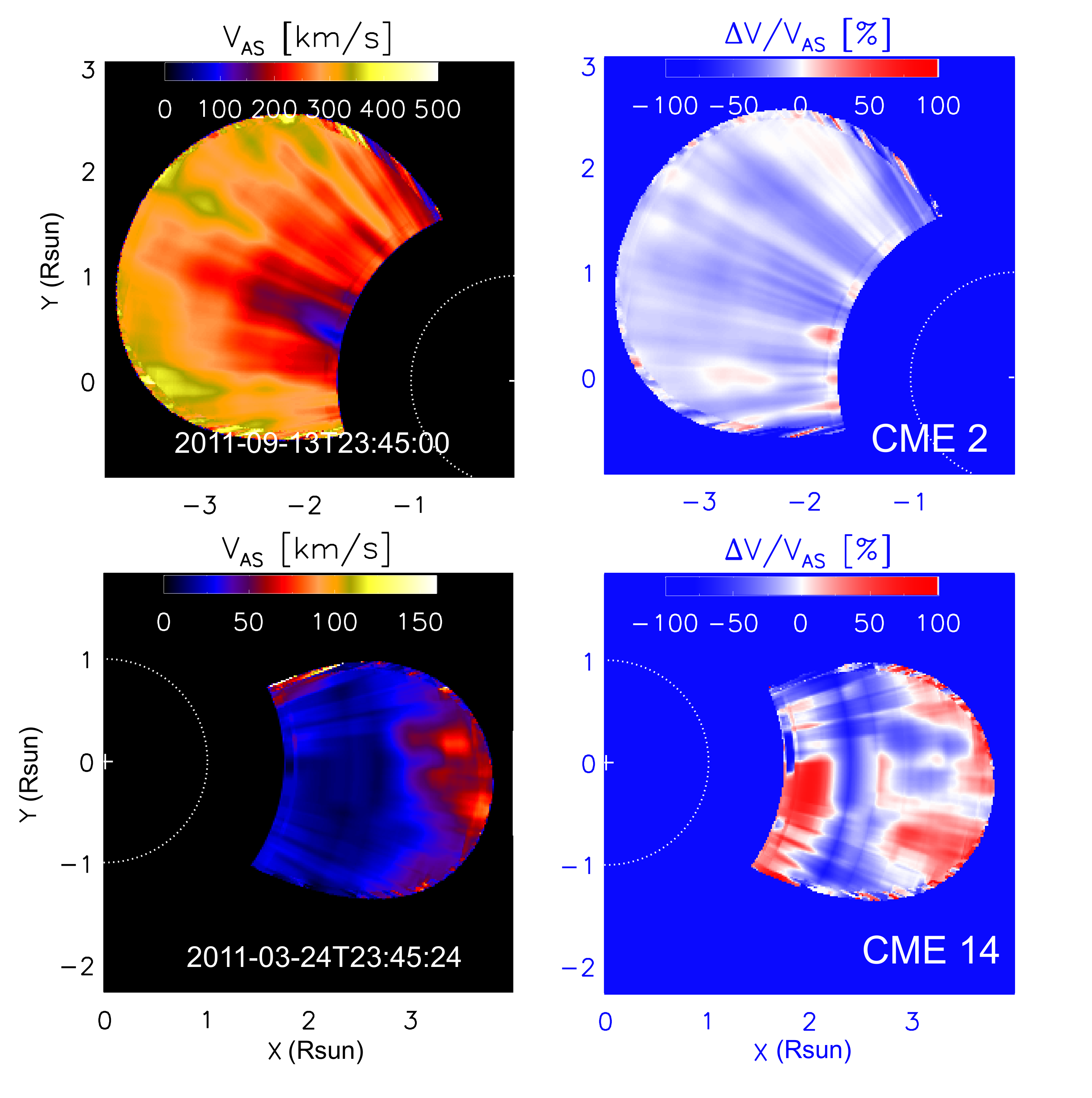}
  
  \caption{Top left: 2D radial speed ($V_{AS}$) distribution of the standard CME 2 by averaging the velocities between the forward step ($V_{FS}$) and backward step ($V_{BS}$) of the cross-correlation method (CCM). Top right: The relative uncertainty ratio of the average of the speed difference $\Delta V$ between $V_{FS}$ and $V_{BS}$ to the average speed $V_{AS}$ for the CME 2. Bottom: The same descriptions as that of top panels but for the stealthy CME 14.}
  \label{fig:eventspeed}
\end{figure*}

\subsection{CME Velocity} ~\label{sec:velocity}
CMEs are usually characterized by significant density inhomogeneities being made by different parts expanding at different speeds. By estimating the CME kinetic energy based on the unidimensional speed determined by tracking the CME brighter front, the result will be generally overestimated \citep{Feng2015a} because the front is generally the region of the CME body expanding much faster. The CCM was built by \citet{Ying2019} to measure the 2D distribution of radial speeds of a CME based on the white-light (WL) coronagraphic images. If we obtain the speed distribution of the CME, we can get more accurate energy distributions of CMEs and study the kinematic features of different parts of the CME.

The CCM requires three consecutive sequences (at $T_{n-1}$, $T_n$, $T_{n+1}$ times) of WL images to measure the CME radial speed at $T_n$ time. These three images are pre-processed into normalized running-difference images (normalized by the previous image) and converted from Cartesian to polar coordinates (the horizontal axis is the radial distance, and the vertical axis is the latitude). As described more in detail by \citet{Ying2019}, through the forward step (FS, by using images at $T_n$ and $T_{n+1}$ times) and backward step (BS, by using images at $T_n$ and $T_{n-1}$ times), we can obtain the displacements of the center pixel of the signal window at $T_n$ pixel by pixel based on the maximal cross-correlation value between two similar signal windows at two different observation times. Subsequently, given the time interval between two consecutive frames, the displacement can be converted to the radial speed at $T_n$ time for each pixel. The average speed, $V_{AS}$, is acquired by averaging the speeds measured by the FS and BS, where $V_{AS}=(V_{BS}+V_{FS})/2$. More details about the CCM can be referred to \citet{Ying2019}.

For the standard CMEs, the signatures in the coronagraphic images are unambiguous. The normalized running-difference images ($\Delta t=5 \rm~minutes$) are enough to track the CME structures via the CCM. Whereas stealthy CMEs are usually too slow and weak to appear obviously in running-images. Thus, we choose a longer time interval ($\Delta t=30 \rm~minutes$) to obtain the normalized difference images, which has been applied to stealth CMEs of \citet{Nitta2017} and \citet{Palmerio2021}. For CMEs 9 and 12, we set $\Delta t=60 \rm~minutes$ due to their extremely weak signal-to-noise ratio. Two examples of the measured average speed ($V_{AS}$) are displayed in the left panels of \autoref{fig:eventspeed}. The average of the speed difference between $V_{FS}$ and $V_{BS}$ is defined as $\Delta V=(V_{FS}-V_{BS})/2$. If $V_{FS}<V_{BS}$, $\Delta V$ will be negative. The right panels of \autoref{fig:eventspeed} show the relative uncertainty ratio of $\Delta V$ to $V_{AS}$. Then, the speed error propagating to the kinetic energy is assumed as $|\Delta V|$, which is always no less than zero.

\subsection{CME Mass} ~\label{sec:mass}

\begin{figure*}[htb]
  \centering
  \includegraphics[width=1\textwidth]{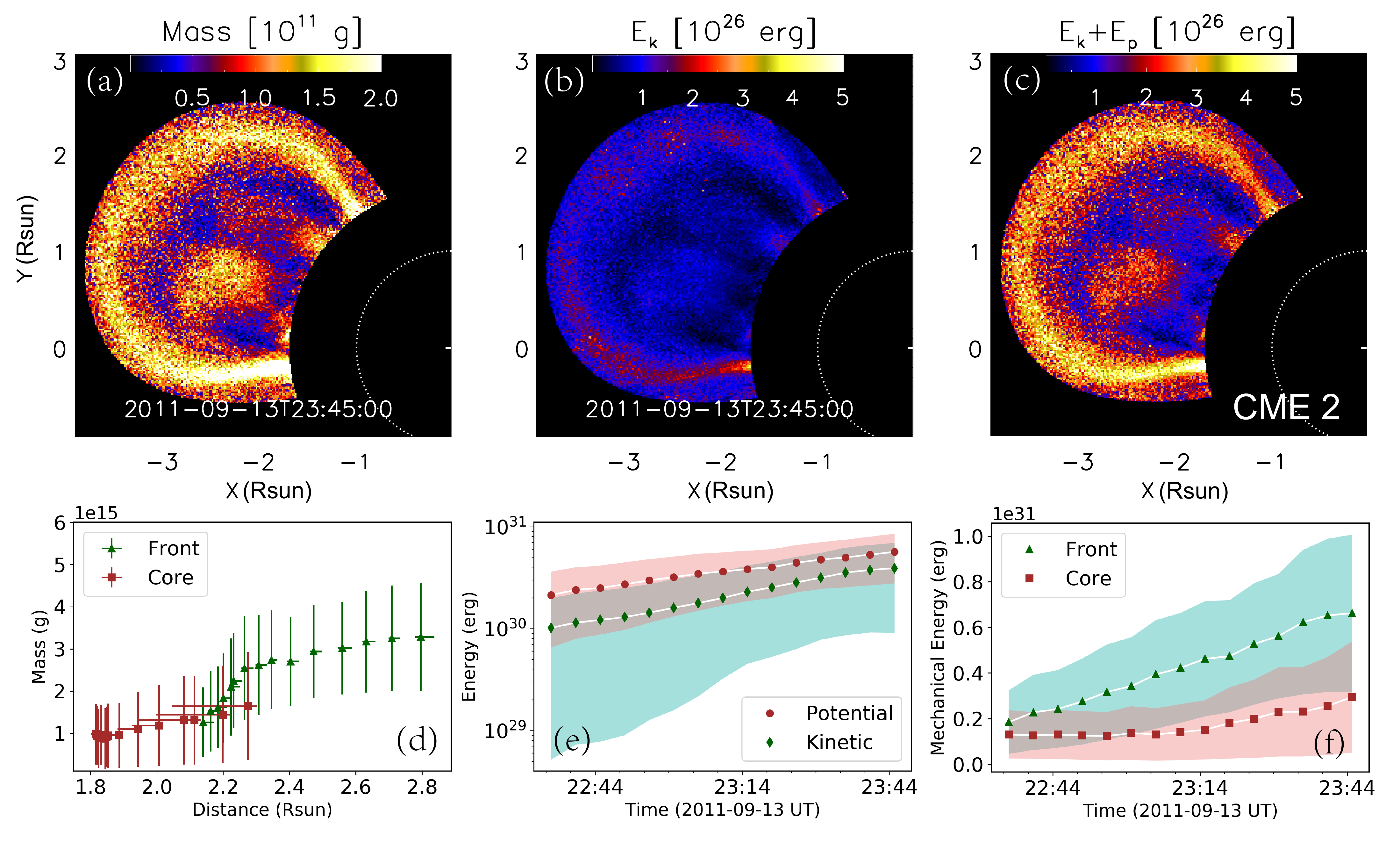}
  
  \caption{The distribution (top) and evolution (bottom) of the mass and energy for the standard CME 2. (a)-(c): The distributions of the CME mass, kinetic energy, and mechanical (kinetic $+$ potential) energy at $\sim$23:45 UT. (d): Mass evolution of the CME core (square) and front (triangle). (e): Evolution of the CME's total kinetic energy (circle) in the radial direction and total potential energy (diamond). (f): Mechanical energy evolution of the CME core (square) and front (triangle), respectively. Shadow areas in panels (e)-(f) denote the upper and lower limits of the energies.}
  \label{fig:eventcme2}
\end{figure*}

\begin{figure*}[htb]
  \centering
  \includegraphics[width=1\textwidth]{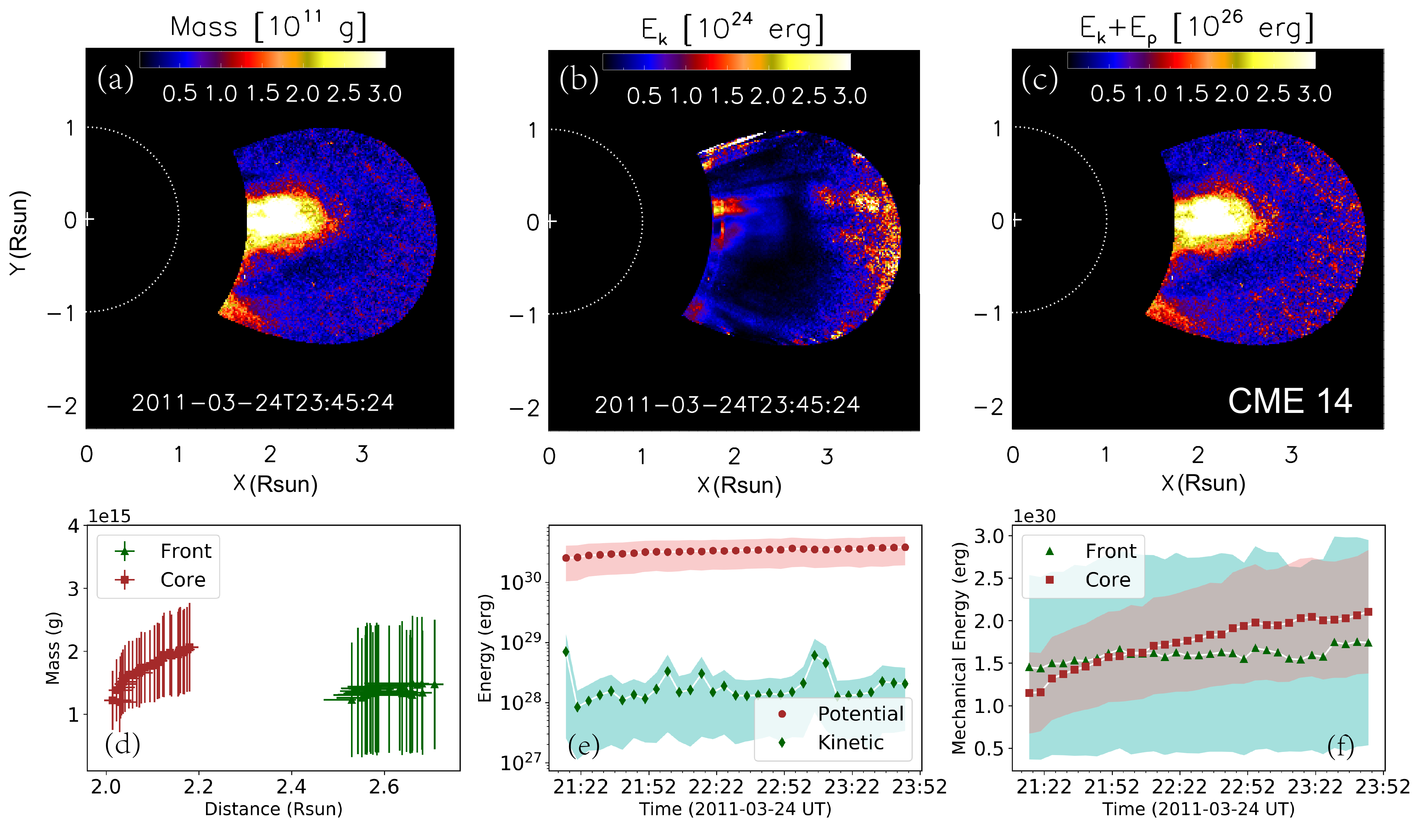}
  
  \caption{The descriptions are the same as that of \autoref{fig:eventcme2} but for the stealthy CME 14.}
  \label{fig:eventcme14}
\end{figure*}

The CME mass can be estimated via WL coronagraphic images yielded by Thomson scattering and linked linearly to the density of the free electrons. We assume all electrons of CMEs lying on the PoS, considering that the Thomson scattering total brightness only decreases by 10\% when the deviation angle of the CME away from the PoS is at 35$\arcdeg$ and keeps almost constant when the angle less than 20$\arcdeg$ \citep{Vourlidas2010}. The significant non-Thomson-scattered contributions (such as stray light and dust scattering) and Thomson-scattered background contributions are generally removed by subtracting the pre-event background image. However, there are bright streamer-like or multi-ray-like structures in the pre-event background image for all events. The subtraction of the pre-event background, including the bright structures, will result in negative values in many pixels of the base-difference image, and the CME mass will be underestimated. According to \citet{Ying2019}, we remove the non-CME signals by subtracting pre-event and 24 hr minimum backgrounds. The 24 hr minimum background refers to the minimum value corresponding to each pixel on the image within 24 hours; thus, some bright streamer- or ray-like structures will be left for mass calculations of CMEs, which will lead to the overestimation of the CME mass. Therefore, the measured mass values based on these two backgrounds can be then considered as lower and upper limits of the CME mass. \autoref{fig:eventcme2} (a) and \autoref{fig:eventcme14} (a) present the mass distributions of CMEs by averaging the upper and lower limits of the CME mass.


\subsection{CME features separation}

Some CMEs in this work show an evident three-part structure, such as CMEs 2 and 14 shown in \autoref{fig:eventimg}, while the remaining CMEs with cores are marked in column 8 of \autoref{tab:01}. The separation of the CME core and front allows us to investigate the dynamic characteristics of different CME parts. According to the method described in \citet{Ying2019}, the possible ``core region'' is first selected by visual inspection based on WL base-difference images after subtracting the 24 hr minimum background. The chosen core region is defined ``by hand'' as a region of interest where the core should be located and will be larger than the actual core region. Subsequently, the selected region is normalized by its maximum intensity, and we define pixels belonging to the CME core as those having WL normalized intensity between 1 and a fraction $f$. We try to ensure that signals which may belong to the CME core are not missed as much as possible during the CME evolution by adjusting the fraction $f$. The optimized fraction $f$ of each CME is listed in the column 9 of \autoref{tab:01}. Then, once we have all pixels belonging to the actual core region, the remaining CME pixels are defined to belong to the ``front$+$void'', which will be simply called ``front'' hereafter. Two examples of the actual core regions are encircled by the white dotted lines, as shown in \autoref{fig:eventimg}. The mass evolution of the separated core (red square) and front (green triangle) are shown in panels (d) for the standard CME in \autoref{fig:eventcme2} and for the stealthy CME in \autoref{fig:eventcme14}.

\subsection{CME Energy}

Given the measured speed and mass distributions of CMEs, we can estimate kinetic, potential, and mechanical energies. The total potential energy, $E_{p}$, of a CME, required to lift the CME plasma from the solar surface to its position, can be calculated by the equation \citep{Vourlidas2000}
\begin{equation}
 	E_{pi}=\int_{R_{\odot}}^{r_i}  \frac{GM_{\odot}m_{i}}{r^2}dr~\rm (erg),
\end{equation}
\begin{equation}
  	E_{p}=\sum_{i=1}^{N} E_{pi}~\rm (erg), 
  \label{eq:eqp}
\end{equation}
where $G$ is the gravitational constant, $M_{\odot}$ is the solar mass, $m_{i}$ is the column mass of the CME for a pixel $i$, and $r_i$ is the radial distance for the pixel $i$ from the solar center, and $N$ is the number of pixels comprising the CME in the 2D image. $E_{pi}$ is the potential energy of the CME for each pixel $i$. $E_{p}$ is the sum of the potential energy $E_{pi}$ of all pixels comprising the CME. The total kinetic energy, $E_{k}$, of the CME in the radial direction can be measured by the formula 
\begin{equation}
  E_{ki}=\frac{1}{2} m_{i}~v_{i}^2~(\rm erg),
\end{equation}
\begin{equation}
  E_{k}=\sum_{i=1}^{N} E_{ki}~(\rm erg),
  \label{eq:eqk}
\end{equation}
 where $v_{i}$ is the radial speed for the pixel $i$ in the PoS measured by the CCM. Similar to the potential energy,  $E_{ki}$ is the kinetic energy of the CME for each pixel. $E_{k}$ is the sum of the kinetic energy $E_{ki}$ of all pixels. In this work, we obtain upper and lower limits of the CME's potential energy and kinetic energy in the radial direction based on the upper and lower limits of the mass and the radial speed, representative of the uncertainties on these parameters as discussed in Sections \ref{sec:velocity} and \ref{sec:mass}. The kinetic energy distributions for each pixel averaging between the upper and lower limits are shown in panels (b) of Figures \ref{fig:eventcme2} and \ref{fig:eventcme14}. Panels (c) show the mechanical energy distributions of the standard CME in \autoref{fig:eventcme2} and stealthy CME in \autoref{fig:eventcme14}. Panels (e) of Figures \ref{fig:eventcme2} and \ref{fig:eventcme14} show the evolution of the total kinetic energy ($E_{k}$, green diamond) in the radial direction and total potential energy ($E_{p}$, red circle) summing from 2D distributions. The total kinetic energy of the stealthy CME is several orders of magnitude lower than the total potential energy. As we mentioned in Section 2, the deviation angle of the CME 15 away from the PoS will lead to an error of about 40\% in the estimation of the kinetic energy. Compared with the total potential energy, the error of the kinetic energy caused by the projection effect has a negligible impact on the mechanical energy. Panels (f) in Figures \ref{fig:eventcme2} and \ref{fig:eventcme14} display the evolution of the mechanical energy of the CME core and front for these two types of CME.

\begin{figure*}[htb]
  \centering
  \includegraphics[width=1.\textwidth]{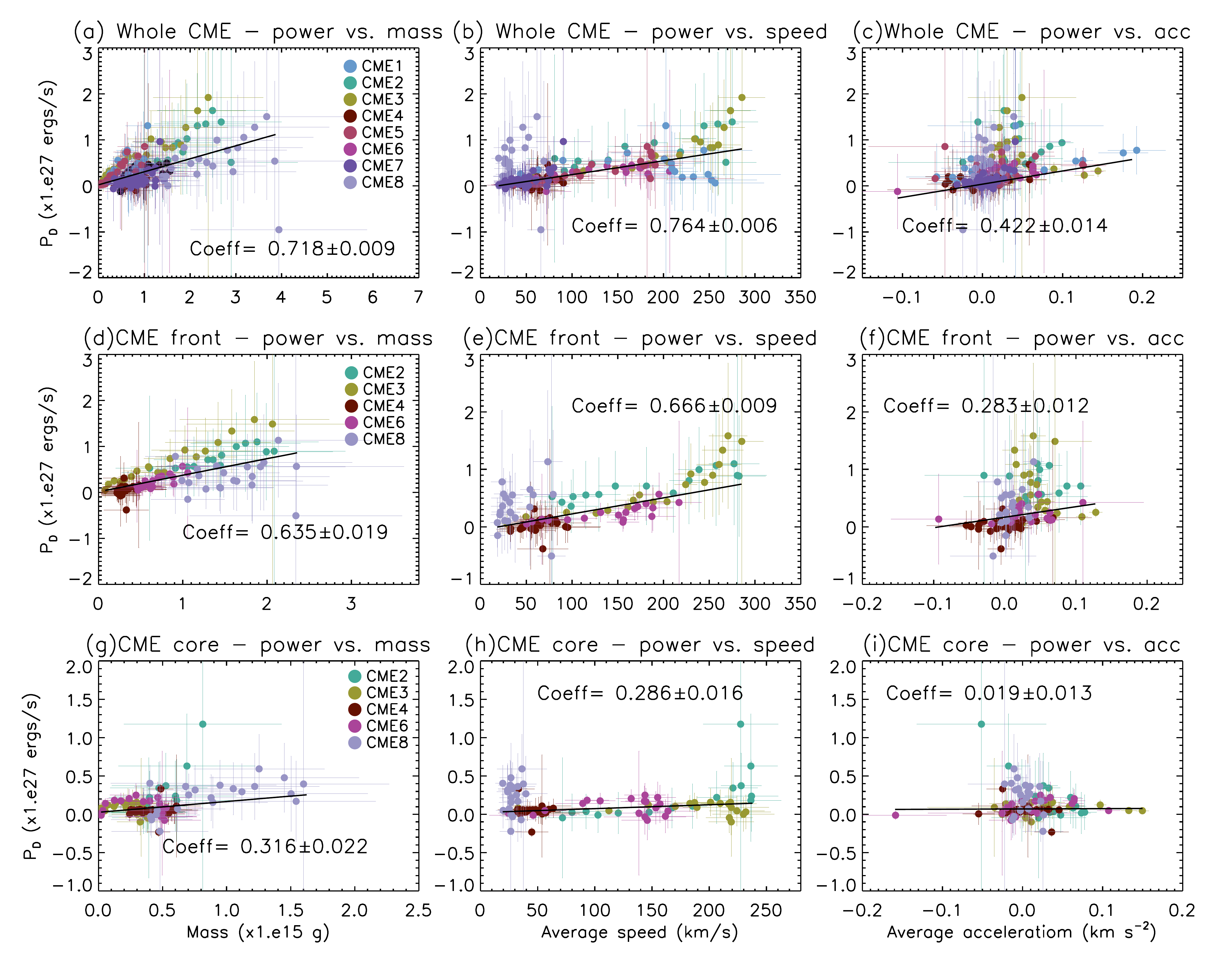}
  
  \caption{Correlations between CME parameters and CME driving powers ($P_{D}$) for standard CMEs. From left to right, panels show the correlation between mass (left), average speed (middle), average acceleration (right) and the driving power. From top to bottom, panels denote the correlations of the whole CMEs (top), CME fronts (middle), and cores (bottom). Some error bars that are out of range in the figure are not displayed.}
  \label{fig:correlation1}
\end{figure*}

\begin{figure*}[htb]
  \centering
  \includegraphics[width=1.\textwidth]{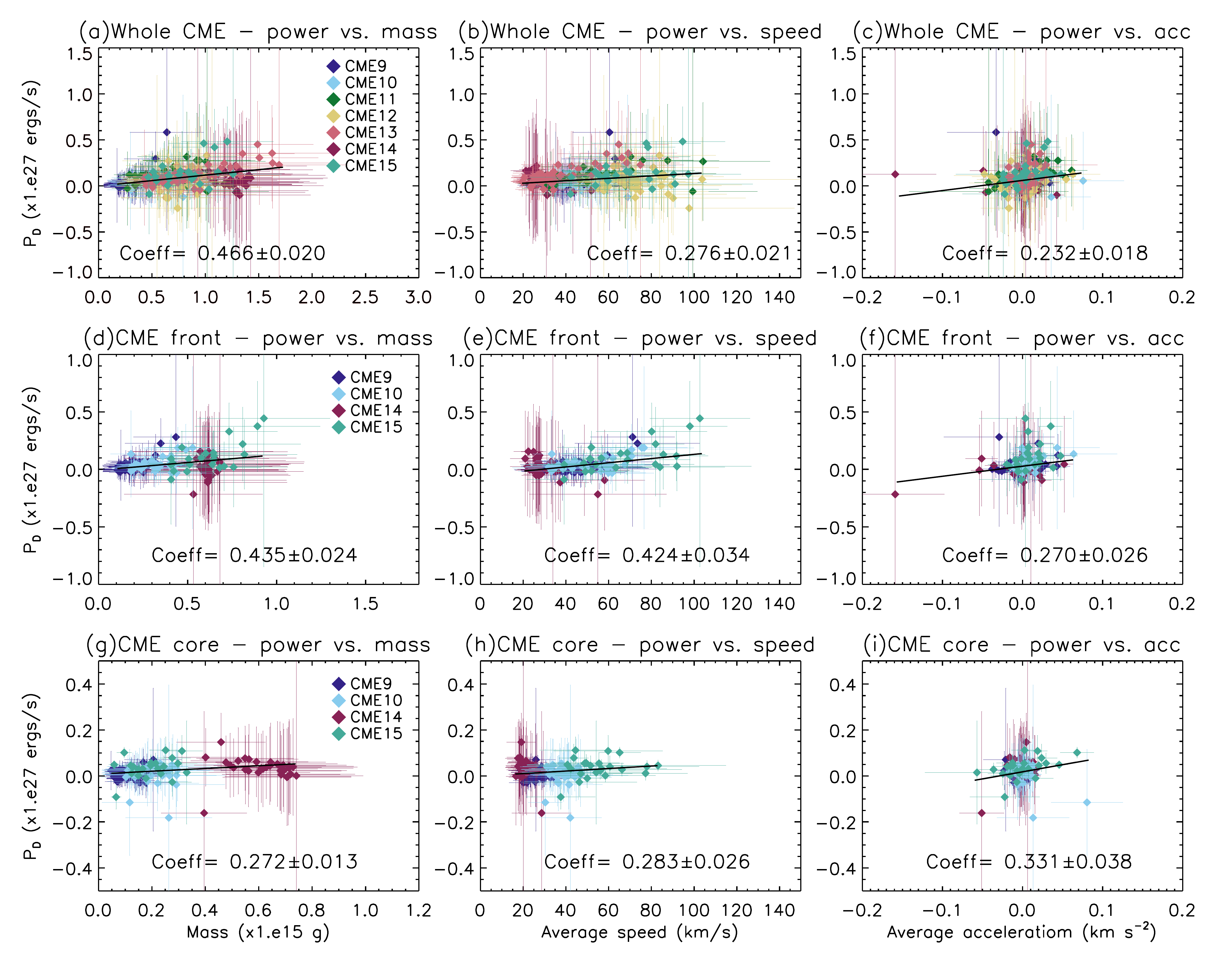}
  
  \caption{Correlations between CME parameters and CME driving powers for stealthy CMEs. Descriptions are the same as that of \autoref{fig:correlation1}.}
  \label{fig:correlation2}
\end{figure*}


\section {Correlations between CME parameters and driving powers} ~\label{sec:result}

The driving power corresponds to work done by some internal or external forces on CMEs \citep{Subramanian2007,Vrsnak2010,Sachdeva2017}. In this section, we investigate the dynamic characteristics of CMEs to derive the driving powers of the whole CMEs and that of different CME structures (core and front) based on mechanical energies. The driving power is assumed to be equal to the mechanical energy rate, which is given by $P_{D}=d(E_{k}+E_{p})/dt$. 
The uncertainty of the driving power is obtained with the error propagation starting from the mass and velocity uncertainties. The correlations between the driving power and different CME parameters are shown in \autoref{fig:correlation1} for the standard CMEs and \autoref{fig:correlation2} for the stealthy CMEs. The CME parameters include the total mass, average speed, and average acceleration. The mass and average speed uncertainties are propagated from the upper and lower limits of the mass and velocity distributions. The error bars of the average accelerations are propagated from the uncertainty of the average speed.

{Results in \autoref{fig:correlation1} (top panels) show that the total driving powers of the standard CMEs are correlated with the mass (correlation coefficient $cc=0.72$, panel a), average speed ($cc=0.76$, panel b), and weakly correlated with the average acceleration ($cc=0.42$, panel c). For some CME events with evident cores and fronts, we separate these two parts to investigate their driving powers. The divided results are shown in the second and third rows in \autoref{fig:correlation1}. Panels (d)-(f) display that the driving powers of the CME fronts have correlations with the mass and speed and weak correlations with the acceleration, similar to the whole CMEs. Compared to the fronts of the standard CMEs, the powers of the CME cores correlate weakly with the mass and speed, and do not correlate with the acceleration (panels g-i).
In this work, the average speed of the whole CME is no more than 300 \kms, as shown in the second columns of \autoref{fig:correlation1} and \autoref{fig:correlation2}. The average potential energy is about five times the average kinetic energy for standard CMEs and thirty times for stealthy CMEs. Thus, the primary contribution of the driving power comes from the potential energy, and the driving power is mainly proportional to the rate of increase in mass. Thus, the slopes in all panels indicate the rate of increase in mass. The larger slopes in the CME front than those in the core imply the primary mass enhancement of the CME coming from the region outside the core (note that we define the front as the region outside the core). The result is reasonable that many studies have revealed the mass enhancement of a CME mainly appearing in the CME front due to the mass pileup from the ambient coronal plasma and the mass supplement from dimming regions in the low corona \citep{Ciaravella2003,Aschwanden2009,Tian2012,Feng2015b}.}

{\autoref{fig:correlation2} shows the same correlations for stealthy CMEs. According to \autoref{fig:correlation2}(a)-(c), the total driving powers of the stealthy CMEs weakly correlate with mass ($cc=0.47$, panel a), average speed ($cc=0.28$, panel b), and acceleration ($cc=0.23$, panel c). The correlations of stealthy CMEs between the total driving power and CME parameters are much smaller than those of standard CMEs (as shown in \autoref{fig:correlation1} a-c). Compared with the fronts of standard CMEs, correlations of stealthy CMEs in the fronts are much weaker with the mass and average speed (second row of \autoref{fig:correlation2}). The smaller slopes in the fronts of stealthy CMEs than those of standard CMEs indicate that the stealthy CMEs have lower mass increase rates. This is consistent with the expectations from the stealthy CMEs having extremely weak low coronal signatures (e.g., coronal dimming) in comparison to the standard CMEs.}


\section {CME driving forces} ~\label{sec:force}
For flux-rope models \citep{Forbes2000,Kliem2006}, the Lorentz force ($F_L$) is usually regarded as the primary force to propel CMEs to a few solar radii \citep{Vrsnak2006, Carley2012, Sachdeva2017}, beyond which the external drag force ($F_D$) may dominate \citep{Cargill2004,Byrne2010,Carley2012}, arising from the dynamic interaction between the CME and surrounding solar wind. Besides, CMEs are obviously affected by the gravitational force ($F_{G}$). Thus, by assuming that no other forces are affecting CMEs during their propagation through the corona (and in particular neglecting the thermal pressure force, because no measurements of plasma temperatures are available), the total force that works on the CME can be expressed as $F_{tot}=F_{L}+F_{D}+F_{G}$. In this section, we quantitatively estimate the possible driving forces of the two types of CMEs during the propagation phase in the FoV of COR1.

\subsection{Drag force}
The aerodynamic drag acting over a magnetic flux-rope expanding in a surrounding environment plasma can be estimated by the formula \citep{Cargill1996,Cargill2004}
\begin{equation}
    F_{D}=-\frac{1}{2}C_D~A_{cme}n_s~m_p(V_{cme}-V_{s})|V_{cme}-V_{s}|.
\end{equation}
where $C_D$ is the drag coefficient, $A_{cme}$ is the CME cross-sectional area, $n_s$ is the solar wind density, $m_p$ is the proton mass, $V_{cme}$ is the CME velocity, and $V_{s}$ is the solar wind speed.

\begin{figure}[htb]
    \centering
    \includegraphics[width=0.7\textwidth]{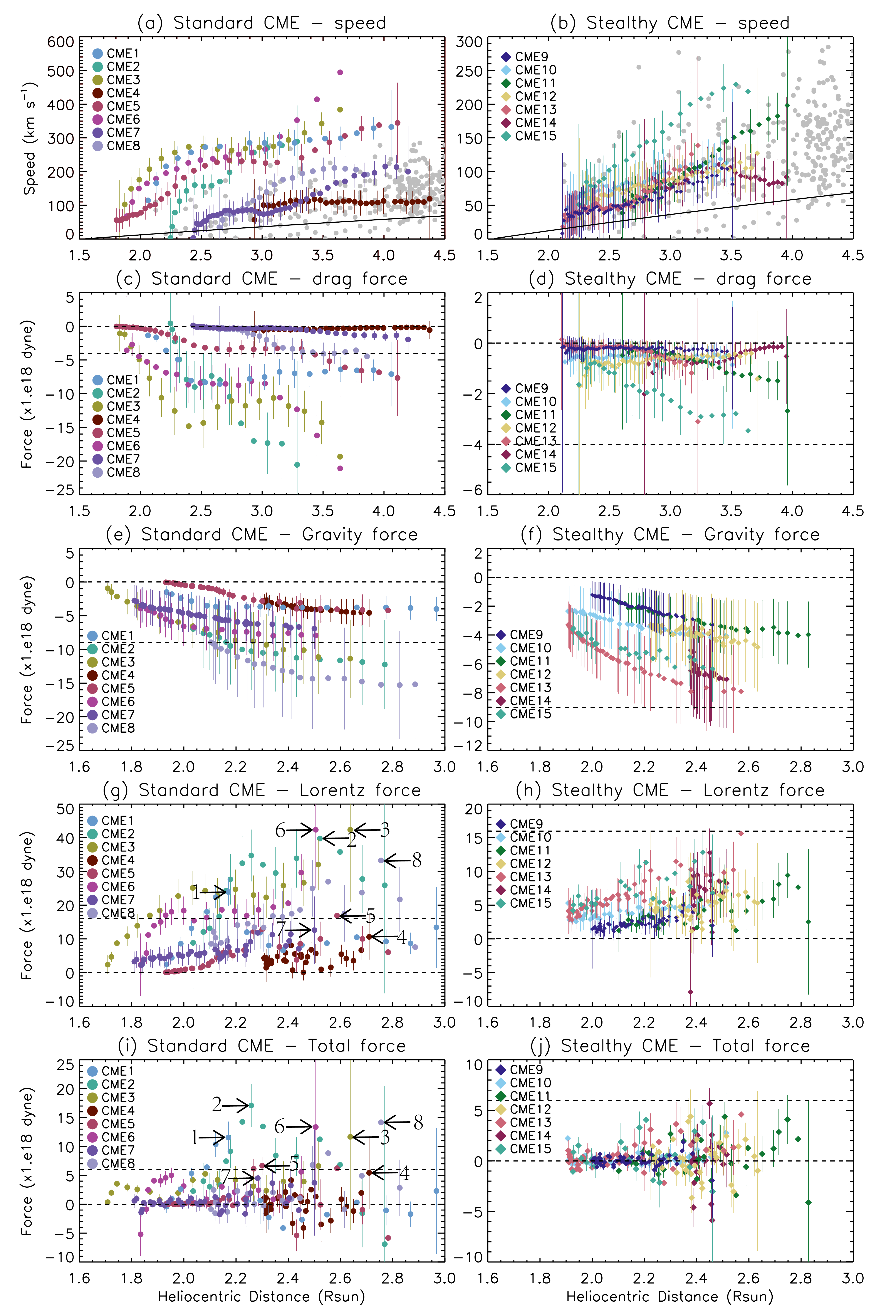}
    \caption{The evolution of the CME parameters along the heliocentric distance for standard (left) and stealthy (right) CMEs. (a)-(b): CME front speeds as a function of CME front distance. The speeds are measured via conventionally tracking bright fronts. A solid line is a fitting profile of the slow solar wind provided by \citet{Bemporad2021}. Gray dots denote slow solar wind speeds measured via streamer blobs from \citet{Sheeley1997}, \citet{Wangym1998}, and \citet{Song2009}. (c)-(d): CME drag force vs. CME front distance. (e)-(f): CME gravitational force vs. CME centroid distance. (g)-(h): CME Lorentz force vs. CME centroid distance. (i)-(j): CME total force vs. CME centroid distance. Arrows in panels (g) and (i) represent the maximum values of the Lorentz force and total force for each standard CME. The horizontal dashed lines in the right panels are the same as those in the corresponding left panels and denote fixed values of the vertical axes.}
    \label{fig:sw}
\end{figure}

Considering that the drag force is expected to mainly act at the CME - solar wind interface (hence on the CME external surface), we use the CME speed measured via conventionally tracking bright fronts instead of the average speed calculated from the 2D velocity map by the CCM. Note that the CME ``front'' in this section refers to the CME leading edge. Besides, we use the heliocentric distance of the front ($R_f$) to derive the slow solar wind speed for each event rather than the centroid distance, calculated by $R_c=\sum m_i~r_i/\sum m_i$, of the whole CME. The derived speed-distance results are shown in \autoref{fig:sw} for standard (panel a) and stealthy (panel b) CMEs. The speed of the CME front has the uncertainty that is propagated from the uncertainty in the distance measurements. Besides, using the latter uncertainty is also propagated to the uncertainties in the solar wind speed, solar wind density, and drag force with Monte Carlo simulations. In these simulations, 100 randomly selected distance values following a Gaussian distribution with $3\sigma=150$ Mm are used at each distance-time data point. We assume that the CME cross-sectional area equals its projected area in the COR1 FoV. The solar wind density, $n_s$, is estimated by \citet{Vrsnak2004}
\begin{equation}
    n_s[10^{8}~cm^{-3}]=\frac{15.45}{R_f^{16}}+\frac{3.16}{R_f^{6}}+\frac{1}{R_f^4}+\frac{0.0033}{R_f^2}.
\end{equation}
As we mentioned above, all CMEs are propagating into the slow solar wind regime. Thus, we use a slow solar wind profile to estimate the drag forces in this work. The profile of the slow solar wind speed is given by \citet{Bemporad2021}
\begin{equation}
    V_s=\frac{74.3R_f-113.4}{0.135R_f+2.61}~(\rm km~s^{-1}),
\end{equation}
where $R_f$ is expressed in $\rm R_{\odot}$. The fitting profile (solid lines in the panels a and b of \autoref{fig:sw}) is derived from observations of the UV Coronagraph Spectrometer \citep[UVCS; ][]{Kohl1995} on board the Solar and Heliospheric Observatory \citep[SOHO; ][]{Domingo1995} and is valid only for $R_f>1.53~\rm R_{\odot}$ \citep{Strachan2002,Noci2007}. Gray dots denote the possible range of slow solar wind speeds measured via streamer blobs from \citet{Sheeley1997}, \citet{Wangym1998}, and \citet{Song2009}. Then, we can estimate the approximate magnitude of the drag force by assuming a constant drag coefficient $C_D=1$ \citep{Cargill2004,Byrne2010,Carley2012}. The results are shown in panels (c) for standard CMEs and (d) for stealthy CMEs in \autoref{fig:sw}. We can find that the absolute values of drag forces acting on the most standard CMEs, except for two CMEs (4 and 7), are larger than those on all stealthy CMEs. The maximum absolute value of drag forces can reach up to $\sim2.1\times10^{19}$ dyne for standard CMEs.
\subsection{Lorentz force}
To estimate the total Lorentz force, we derive the total force acting on the whole CME first, according to the basic physical formula $F_{tot}=ma$, where $m$ denotes the mass of the entire CME, and $a$ is the average acceleration via the average speed derived by the CCM. Then, the Lorentz force can be computed by $F_{L}=F_{tot}-F_{D}-F_{G}$. Note that the total Lorentz force consists of the outward Lorentz self-force (or hoop force) due to the current in the flux rope and the compressing Lorentz force due to a combination of the external poloidal magnetic fields in the background and the current inside the flux rope \citep{Chen1989, Chen1996, Kliem2006}. In this work, we ignore the plasma pressure gradient, which is usually directed away from the Sun. We derive $F_{G}$ by the formula, $F_{G}=\sum_{i=1}^{N}\frac{G M_{\odot} m_i}{r_i^2}$, which depends only on the mass and heliocentric position of the CME. Because the direction of $F_{G}$ always points to the solar center, the results are negative compared to the propagation direction of the CME. The results of the gravitational forces, Lorentz forces, and total forces of the standard and stealthy CMEs are shown in \autoref{fig:sw}. It is noted that the distance in panels (e)-(j) denotes the centroid position of the CME, given the gravity, Lorentz, and total forces work on the entire CME. The combination of the drag and gravitational forces is comparable with and even can be larger than the Lorentz force for both standard and stealthy CMEs in the inner corona, considering the evolution of the total force as shown in panels (i) and (j) of \autoref{fig:sw}. The negative values of total forces reflect the deceleration of the whole CMEs. The peak value of the Lorentz and total forces for each standard event are marked by black arrows in panels (g) and (i). Compared to the Lorentz forces of the standard CMEs, several peak positions of the total forces change a lot (such as CMEs 2, 5, 7) along the heliocentric distance due to the impact of the drag and gravitational forces. The maximum absolute value of gravitational forces is $\sim 1.5\times 10^{19}$ dyne for standard CMEs. The maximum Lorentz force of standard CMEs can reach up to $\sim 4.3\times10^{19}$ dyne, which is similar to the magnitude of the Lorentz force acting on the CMEs in \citet{Sachdeva2017} and \citet{Carley2012}. Note that the Lorentz forces of most standard CMEs go through the increase and decrease patterns between 1.7 and 3 \rsun, except for CMEs 4 and 7. Compared with the standard CMEs, the maximum Lorentz force of stealthy CMEs is small and no more than $\sim 1.6\times10^{19}$ dyne. The evolution of the Lorentz force in stealthy CMEs is more gradual without the clear increase and decrease pattern. Besides, an interesting difference between the front speed evolution of standard and stealthy CMEs appears in panels (a) and (b) of \autoref{fig:sw}. Most of the standard CMEs present a ``shoulder" in the evolution of the speed within the distance of 2-3 \rsun, while the stealthy CMEs are more gradual. The speed difference can manifest that the effects of drag, Lorentz, and total forces on CME are more pronounced on standard CMEs than stealthy CMEs. On the other hand, comparing the drag, gravity, Lorentz, and total forces between the stealthy CMEs and standard CMEs (such as CMEs 4 and 7), these parameters appear to have a similar evolution tendency. 


\section {Discussion and conclusions} ~\label{sec:conclusion}
We analyze the kinematics of stealthy and standard CMEs in the FoV of COR1 based on the cross-correlation method developed by \citet{Ying2019}. This method can provide a 2D map of the CME radial speed, and a more accurate estimate of energy distribution, thus allowing to investigate kinematic features of the CME core and front separately. We studied the possible correlations between CME driving powers and their physical parameters (including the total mass, average speed, and average acceleration) and estimated the CME driving forces to explore the dynamic similarities and differences in the dynamical behaviors of stealthy and standard CMEs. 

Based on the correlation results between the driving power and CME parameters, we find that the driving powers of the front and the whole part of standard CMEs show strong correlations with the mass and speed. The slope indicating the mass increase rate is higher in the CME front than in the core, which implies that the primary mass enhancement comes from the front. For the stealthy CMEs, the driving powers only weakly correlate with the mass and speed. Compared with the standard CME front, the slopes in the fronts of stealthy CMEs are much lower, demonstrating a smaller mass pile-up due to the interaction with the ambient coronal plasma and mass supply from the low corona. This result is consistent with the view that the stealthy CMEs have a weaker interaction with the background corona and solar wind and more unnoticeable low coronal features that may indicate mass supply.

From the perspective of the driving force, standard CMEs are likely propelled by the combined action of the Lorentz, gravity, and drag forces. The Lorentz force acting on the CME is dominant, however, the impact of the gravitational force and the drag force due to the ambient solar wind is non-negligible, as well, at least in the FoV of COR1. We find that the Lorentz force undergoes an increase and then a decrease within the range of 1.7 and 3 \rsun, except for two events (CMEs 4 and 7). The maximum of the Lorentz force is around $4.3\times10^{19}$ dyne. Compared with the standard CMEs, the dynamic effects of the Lorentz and drag forces are significantly smaller for stealthy CMEs near the Sun. However, the gravitational forces working on the stealthy CMEs are similar to those on the standard CMEs, because the gravitational forces only depend on the CME mass and heliocentric position. Furthermore, for stealthy CMEs, the evolution of the Lorentz force as a function of the heliocentric distance is more gradual, without an obvious increase and decrease pattern. Comparing the drag, gravitational, Lorentz, and total forces between the stealthy CMEs and two standard CMEs, the stealthy CMEs perform similar dynamical properties to these slow standard CMEs, which suggests an intersection of these two types of CMEs in dynamics.

We find that a flux rope modelled by \citet{Chen1996} shows the increase and decrease pattern in the Lorentz force profile due to a variation in the total toroidal current \citep[Figure 7 in ][]{Chen1996}. The Lorentz force in this model mainly comes from the internal hoop force. On the other hand, a gradually increased profile of the Lorentz force acting on the CME is shown in a 2.5-dimensional magnetohydrodynamics simulation \citep[in Figure 8 of][]{Zhao2017ApJ}. \citet{Zhao2017ApJ} modelled a flux rope triggered by the photospheric converging motion. Thus, the Lorentz force is more likely an external Lorentz pressure force. A comparison with the Lorentz force profiles shown in these two papers seems to suggest that the ``standard CMEs" are accelerated mainly by the internal hoop Lorentz force, while the ``stealthy CMEs" are accelerated primarily by the external Lorentz pressure force. It is noted that the eruption in \citet{Zhao2017ApJ} only covers about 40 minutes, while the main variation of the Lorentz force shown in Figure 7 of \citet{Chen1996} occurs over about 70-80 minutes. If the simulation of \citet{Zhao2017ApJ} can continue, the Lorentz force would likely decrease later on, similar to that in \citet{Chen1996}.

In this work, we use a constant drag coefficient of unity to estimate the drag force acting on the CMEs. \citet{Subramanian2012} emphasized the relationship between the drag coefficient and the Reynolds number and obtained an analytical model for the drag coefficient through a microphysical prescription for viscosity in turbulent solar wind. The obtained drag coefficient decreases monotonically with the heliocentric distance. The value of the drag coefficient in \citet{Subramanian2012} is around 0.7 at the heliocentric distance of $\sim 5$ \rsun. Thus, the drag coefficient of unity in the inner corona seems reasonable compared with the result obtained by \citet{Subramanian2012}.  

As it is usually agreed, the Lorentz force is responsible for initiating CMEs \citep{Forbes2000} and is assumed to be dominant in their dynamics within a distance of a few solar radii \citep{Vrsnak2006}. \citet{Subramanian2007} supported the idea that the drag force of the solar wind can be ignored in the FoV of LASCO due to the poor correlation between the CME driving force and the cross-sectional area of CMEs. On the other hand, \citet{Byrne2010} and \citet{Carley2012} demonstrated that the drag force acting on a slow CME could be dominant beyond a distance of $\sim7$ \rsun. In this work, we find that the drag and gravitational forces are comparable with the Lorentz force for the slow CMEs even closer to the Sun. Thus, the effect of the drag and gravity should also be taken into account in modeling the CME propagation from the inner corona to improve the prediction of its arrival. 

Before concluding, we point out that in this work, because of the lack of plasma temperature measurements inside CMEs, we neglected the possible role played of thermal pressure force in the dynamic of CMEs. Nevertheless, thermal pressure force may be important for the conversion of energy into momentum in the early acceleration phase \citep{Wu2004, Liu2008}, as well as during the interplanetary propagation phase as the internal driver of the CME expansion \citep{Wang2009, Mishra2018}. For these reasons, in the future we plan to extend the analysis presented here to events being now observed by the Metis coronagraph on-board Solar Orbiter mission \citep{Antonucci2020}, as well as to those that will be observed by the Lyman-alpha Solar Telescope instrument \citep{Li2019,Feng2019} on-board the Chinese Advanced Space-based Solar Observatory mission \citep{Gan2019}. In fact, as it was recently demonstrated by \citet{Bemporad2022}, visible light and UV Lyman$-\alpha$ coronagraphic observations of CMEs can be combined to derive the 2D distribution of thermal energy inside CMEs. Applying similar methods to the data provided by the above instruments will thus allow us to extend the analysis presented here.

\acknowledgements
 STEREO is a project of NASA. The SECCHI data used here were produced by an international consortium of the Naval Research Laboratory (USA), Lockheed Martin Solar and Astrophysics Lab (USA), NASA Goddard Space Flight Center (USA), Rutherford Appleton Laboratory (UK), University of Birmingham (UK), Max-Planck-Institut for Solar System Research (Germany), Centre Spatiale de Li{\`e}ge (Belgium), Institut d'Optique Th{\'e}orique et Applique{\'e} (France), Institut d'Astrophysique Spatiale (France). This work is supported by NSFC (grant Nos. U1731241, 11921003, 11973012), CAS Strategic Pioneer Program on Space Science (grant Nos. XDA15052200, XDA15320103, and XDA15320301), the mobility program (M-0068) of the Sino-German Science Center, and the National Key R\&D Program of China (2018YFA0404200). B.Y. acknowledges the CAS Special Research Assistant Project for financial support. NVN acknowledges support from NASA grants NNX17AB73G and 80NSSC20K0217. 





\bibliography{refs}
\end{document}